\documentclass[12pt]{article}
\usepackage{epsf}
\title{ {\bf
Scalar dark matter-Higgs coupling in the case of electroweak
symmetry breaking driven by unparticle }}
\author{\vspace{1cm}\\
        {\bf E. O. Iltan}
        \thanks{E-mail address:
        eiltan@newton.physics.metu.edu.tr}
\\
Physics Department, Middle East Technical University \\
        Ankara, Turkey \\
        }

\date{}

\begin{document}
\setlength{\baselineskip}{24pt}
\maketitle
\setlength{\baselineskip}{7mm}
\begin{abstract}
We study the possible annihilation cross section of scalar dark
matter and its coupling $\lambda_D$ to the standard model Higgs in
the case of the electroweak symmetry breaking driven by
unparticle. Here the annihilation process occurs with the help of
three intermediate scalars which appear after the mixing. By
respecting the annihilation rate which is compatible with the
current relic density we predict the tree level coupling
$\lambda_D$. We observe that the unparticle scaling $d_u$ plays a
considerable role on the annihilation process and, therefore, on
the coupling $\lambda_D$.
\end{abstract}
\thispagestyle{empty}
\newpage
\setcounter{page}{1}
%%%
%%%
%\section{Introduction}
%%
%%
Research aims to understand the nature of the dark matter reaches
great interest since it contributes almost $23\%$ of present
Universe \cite{JungmanG, JungmanG2, KomatsuE}. The existence of
dark matter can not be explained in the framework of the standard
model (SM) and, therefore, one needs to go beyond. There exist a
number of dark matter candidates in various scenarios such as
Supersymmetry, minimal Universal Extra Dimensions and Little Higgs
with Tparity. It is believed that a large amount of dark matter is
made of cold relics and they are in the class nonrelativistic cold
dark matter. Weakly Interacting Massive Particles (WIMPs) are
among such dark matter candidates with masses in the range $10$
GeV- a few TeV. They are stable, interacting only through weak and
gravitational interactions and disappearing by pair annihilation
(see for example \cite{DEramoF, WanLeiGuo} for further
discussion). From the theoretical point of view a chosen discrete
symmetry drives the stability in many scenarios mentioned above .

In the present work we introduce an additional scalar SM singlet
field $\phi_D$, called as darkon, which was considered first by
Silveria \cite{SilveiraV} and studied by several authors
\cite{HolzE, McDonaldJ, PattB, BertolamiO, DavoudiaslG, HeXG}, and
take the lagrangian obeying the $Z_2$ symmetry $\phi_D\rightarrow
-\phi_D$
\begin{eqnarray}
L_D=\frac{1}{2}\,\partial_\mu\,\phi_D\,\,\partial^\mu\,\phi_D-
\frac{\lambda}{4}\,\phi_D^4-\frac{1}{2}\,m_0^2\,\phi_D^2-\lambda_D\,
\phi_D^2\,(\Phi_1^\dagger\, \Phi_1)\, ,\label{Vint}
\end{eqnarray}
where $\Phi_1$ is the SM Higgs field and $\phi_D$ has no vacuum
expectation value. The $Z_2$ symmetry considered ensures that the
darkon field can appear as pairs and they are stable in the sense
that they do not decay into any other SM particles. They can
disappear by annihilating as pairs into SM particles with the help
of the exchange particle(s). Now, we will introduce an electroweak
symmetry  breaking mechanism and study its effect on the Darkon
annihilation cross section.

A possible hidden sector beyond the SM can be among the candidates
to explain this breaking. Such a hidden sector has been proposed
by Georgi \cite{Georgi1,Georgi2} as a hypothetical scale invariant
one with non-trivial infrared fixed point. It is beyond the SM at
high energy level and it comes out as new degrees of freedom,
called unparticles, being massless and having non integral scaling
dimension $d_u$. The interaction(s) of unparticle with the SM
field(s) in the low energy level is defined by the effective
lagrangian (see for example \cite{SChen}). The possibility of the
electroweak symmetry breaking due to the mixing between the
unparticle and the Higgs boson has been introduced in \cite{JPLee}
(see also \cite{Kikuchi2}). In \cite{JPLee} the idea was based on
the interaction of the SM scalar sector with the unparticle
operator in the form $\lambda\,(\Phi^\dagger\,\Phi)\,O_U$ where
$\Phi$ is the SM scalar and $O_U$ is the unparticle operator with
mass dimension $d_u$ (see \cite{Rajaraman, ADelgado, ADelgado2,
Feng, Kikuchi, RZwicky}). By using the fact that unparticles look
like a number of $d_u$ massless particles with mass dimension one
and, therefore,  the operator $O_u$ can be considered in the form
of $(\phi^*\,\phi)^\frac{d_u}{2}$, the interaction term
\begin{eqnarray}
V\sim\lambda\,(\Phi^\dagger\,\Phi)\,(\phi^*\,\phi)^\frac{d_u}{2}\,,
\label{Vint2}
\end{eqnarray}
is induced and it drives the electroweak symmetry breaking in the
tree level \cite{JPLee}\footnote{See \cite{Weinberg} for the
necessity of the radiative corrections for the electroweak
symmetry breaking from hidden sector due to the interaction in the
form $\lambda\,(\Phi^\dagger\,\Phi)\,\phi^*\,\phi$.}. Recently, in
\cite{Iltanmuonanogunp}, this idea has been applied to the
extended scalar sector  which was obtained by introducing a shadow
Higgs one, the complex scalar $\phi_2$ in addition to the SM
Higgs\footnote{Here the $U(1)_s$ invariant Lagrangian including
the shadow sector and the SM one reads
\begin{eqnarray}
L=L_{SM}-\frac{1}{4}\,X^{\mu\nu}\,X_{\mu\nu}+|\Big(\partial_\mu-
\frac{1}{2}\,g_s\,X_\mu\Big)\,\phi_2|^2- V(\Phi_1, \phi_2,
\phi)\nonumber
\end{eqnarray}
where $g_s$ is the gauge coupling of $U(1)_s$ (see
\cite{WFChang}).}. This choice leads to a richer scalar spectrum
which appears as three scalars, the SM Higgs $h_I$, $h_{II}$ and
the heavy $h_{III}$, after the mixing mechanism (see Appendix for
a brief explanation of the toy model used.). These three scalars
are the exchange particles in our analysis and we have the
annihilation process $\phi_D\, \phi_D\rightarrow h_I\,
(h_{II},\,h_{III})\rightarrow X_{SM}$. The total averaging
annihilation rate of $\phi_D\,\phi_D$ reads
\begin{eqnarray}
<\sigma\,v_r>&=&\frac{8\,\lambda_D^2\,
(n_0^2\,\rho_0^2)}{2\,m_D}\,\Bigg|\,\frac{c_\alpha^2}{
(4\,m_D^2-m_{I}^2)+i\,m_{I}\,\Gamma_I}+
\frac{s_\alpha^2\,c_\eta^2}{ (4\,m_D^2-m_{II}^2)+i\,
m_{II}\,\Gamma_{II}}\nonumber
\\ &+& \frac{s_\alpha^2\,s_\eta^2}{
(4\,m_D^2-m_{III}^2)+i\,m_{III}\,\Gamma_{III}}\,\Bigg |^2\,
\Gamma(\tilde{h}\rightarrow X_{SM})\, ,  \label{sigmavr}
\end{eqnarray}
where $\Gamma(\tilde{h}\rightarrow
X_{SM})=\sum_i\,\Gamma(\tilde{h}\rightarrow X_{i\,SM})$ with
virtual Higgs $\tilde{h}$ having mass $2\,m_D$ (see \cite{BirdC,
BirdC2}). Here $v_r$ is the average relative speed of two
darkons\footnote{Here the assumption is that the speed of the dark
matter $\phi_D$ is small enough to have the approximated result
(see for example \cite{HeXG})}, $v_r=\frac{2\,p_{CM}}{m_D}$ with
center of mass momentum $p_{CM}$. The total annihilation rate can
be restricted by using the present dark matter (DM) abundance. The
WMAP collaboration \cite{WAMP} provides a precise determination of
the present DM abundance (at two sigma level) as
\begin{eqnarray}
\Omega\,h^2=0.111\pm 0.018 \, . \label{RelDens}
\end{eqnarray}
Finally, by using the expression which connects the annihilation
cross section to the relic density
\begin{eqnarray}
\Omega\,h^2=\frac{x_f\,10^{-11}\,GeV^{-2}}{< \sigma\,v_r>} \,,
\end{eqnarray}
with $x_f\sim 25$ (see for example \cite{JungmanG2, HeXG,
ServantG, Gopalakrishna, Gopalakrishna2}) one gets the bounds
\begin{eqnarray}
< \sigma\,v_r>=0.8\pm 0.1 \, pb \,,\nonumber
\end{eqnarray}
which of the order of $(1-2)\times 10^{-9}\, GeV^{-2}$. This is
the case that s-wave annihilation is dominant (see \cite{KolbEW}
for details.).
%
%%%
%%%
\\ \\
{\Large \textbf{Discussion}}
\\ \\
%\section{Discussion}
%
In the present work  we extend the scalar sector by considering a
shadow Higgs one with complex scalar and in order to achieve the
electroweak symmetry breaking at tree level we assume that the
unparticle sector, proposed by Georgi \cite{Georgi1}, couples to
both scalars. Furthermore we introduce so called darkon field,
which is a SM singlet with vanishing vacuum expectation value and
it couples to the SM Higgs doublet, with coupling $\lambda_D$.
After the symmetry breaking considered in our toy model the tree
level interaction $DDh$ appears with strength $v_0\lambda_D$ and
this coupling is responsible for the annihilation cross section
which agrees with the present observed dark matter relic density
(eq.(\ref{RelDens})).

Here we study the dependence of coupling $\lambda_D$ to the
parameters of the model used, the Darkon mass, the scale dimension
$d_u$ and the parameter $s_0$. In our calculations we take the
Darkon mass in the range of $10\leq m_D\leq 80$ and use the
central value of the annihilation cross section, namely $ <
\sigma\,v_r>=0.8\, pb$. Notice that in the toy model we consider
there exist three intermediate scalars which appear after the
mixing and they drive the annihilation process with different
couplings.

\begin{figure}[htb]
\vskip -3.0truein \centering \epsfxsize=6.8in
\leavevmode\epsffile{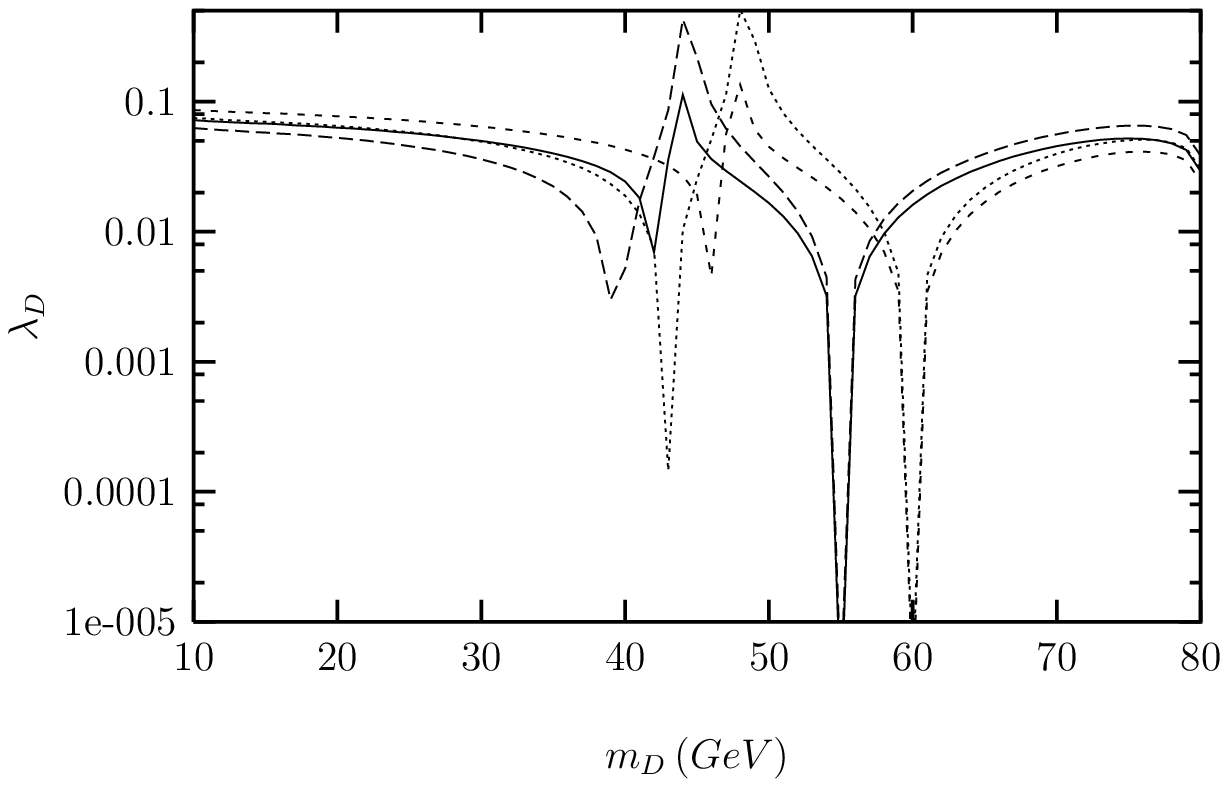} \vskip -3.0truein
\caption[]{$\lambda_D$ as a function of $m_D$ for $d_u=1.1$. Here
the solid-long dashed-dashed-dotted line represents $\lambda_D$
for $m_I=110\,GeV$, $s_0=0.1$-$m_I=110\,GeV$,
$s_0=0.5$-$m_I=120\,GeV$, $s_0=0.1$-$m_I=120\,GeV$, $s_0=0.5$. }
\label{LambDmsdu11b}
\end{figure}

\begin{figure}[htb]
\vskip -3.0truein \centering \epsfxsize=6.8in
\leavevmode\epsffile{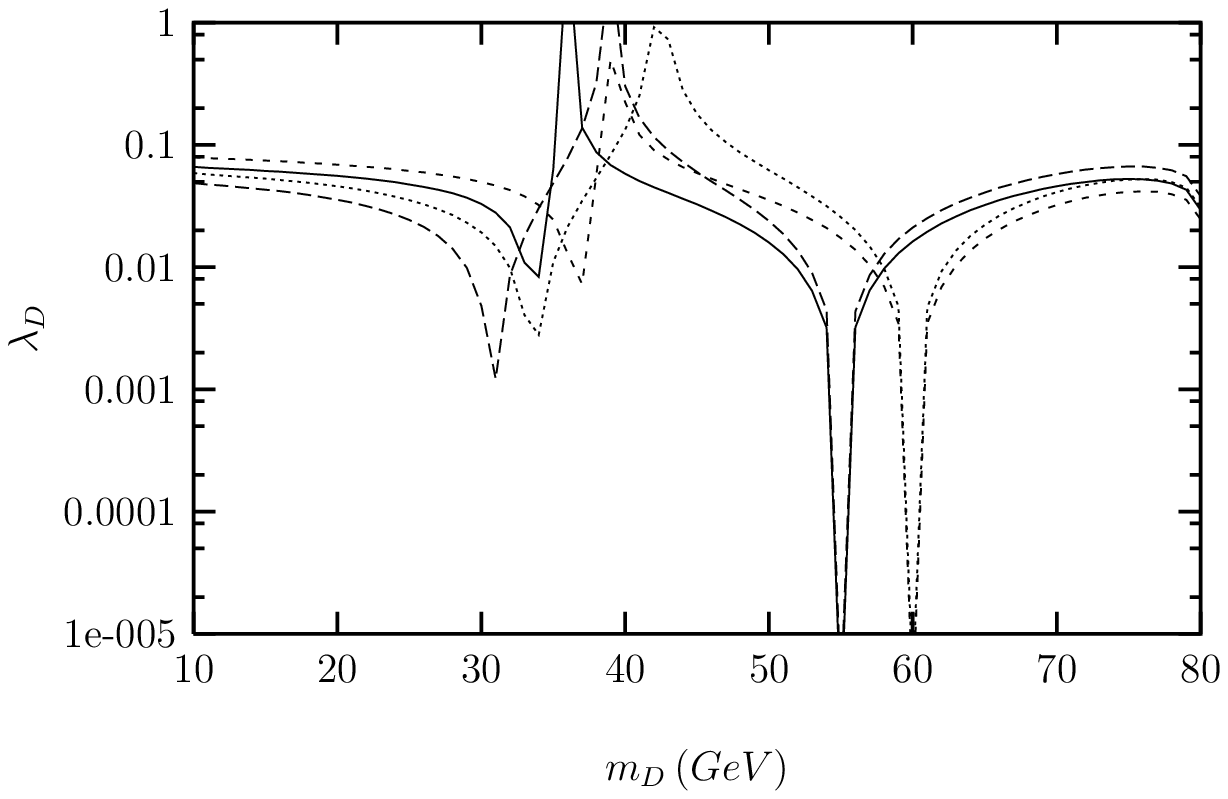} \vskip -3.0truein
\caption[]{The same as Fig. \ref{LambDmsdu11b} but for $d_u=1.5$.}
\label{LambDmsdu15b}
\end{figure}

In Fig.\ref{LambDmsdu11b} we plot $m_D$ dependence of $\lambda_D$
for $d_u=1.1$. Here the solid-long dashed-dashed-dotted line
represents $\lambda_D$ for $m_I=110\,GeV$,
$s_0=0.1$-$m_I=110\,GeV$, $s_0=0.5$-$m_I=120\,GeV$,
$s_0=0.1$-$m_I=120\,GeV$, $s_0=0.5$. $\lambda_D$ is of the order
of $0.1$ for the small values of $m_D$, $m_D \leq 30\,GeV$, and
for the range $m_D \geq 70\,GeV$, in the case that the mass
$m_{I}$ is restricted to $m_{I}=110\, GeV$ and $120\,GeV$. In the
intermediate region $\lambda_D$ drops and increases drastically.
Since there exist three intermediate particles, at the values of
$m_D$ which satisfies the equalities $m_D=\frac{m_i}{2}$ (
$i=I,II,III$), $\lambda_D$ decreases to reach the appropriate
annihilation cross section which is compatible with the current
relic density. In the figure we have two small values of
$\lambda_D$ for each set of $m_D,\,d_u$ and $s_0$ due to resonant
annihilations and the third one is out of the chosen $m_D$ range
which does not include the mass values of the order of
$\frac{m_{III}}{2}$. On the other hand $\lambda_D$ increases up to
the values $0.5$ due to the effects of possible interferences of
intermediate scalar propagators. The figure shows that increasing
mass of $m_{I}$ result in a shift of $\lambda_D$ curve and, for
its increasing values, $\lambda_D$ increases (decreases) for light
(heavy) Darkon in order to satisfy the observed relic abundance.
Fig.\ref{LambDmsdu15b} is the same as Fig.\ref{LambDmsdu11b} but
for $d_u=1.5$. It is observed that there is an considerable
enhancement in the coupling $\lambda_D$ for the some intermediate
values of $m_D$ and first suppressed value(s) of $\lambda_D$
appears for lighter $m_D$ compared to case of $d_u=1.1$. This is
due to the fact that the mass $m_{II}$ becomes lighter with the
increasing values of $d_u$ and the resonant annihilation occurs
for a lighter Darkon mass.

Now we study $d_u$ and $s_0$ dependence of the coupling
$\lambda_D$ to understand their effect on the annihilation rate
more clear.

\begin{figure}[htb]
\vskip -3.0truein \centering \epsfxsize=6.8in
\leavevmode\epsffile{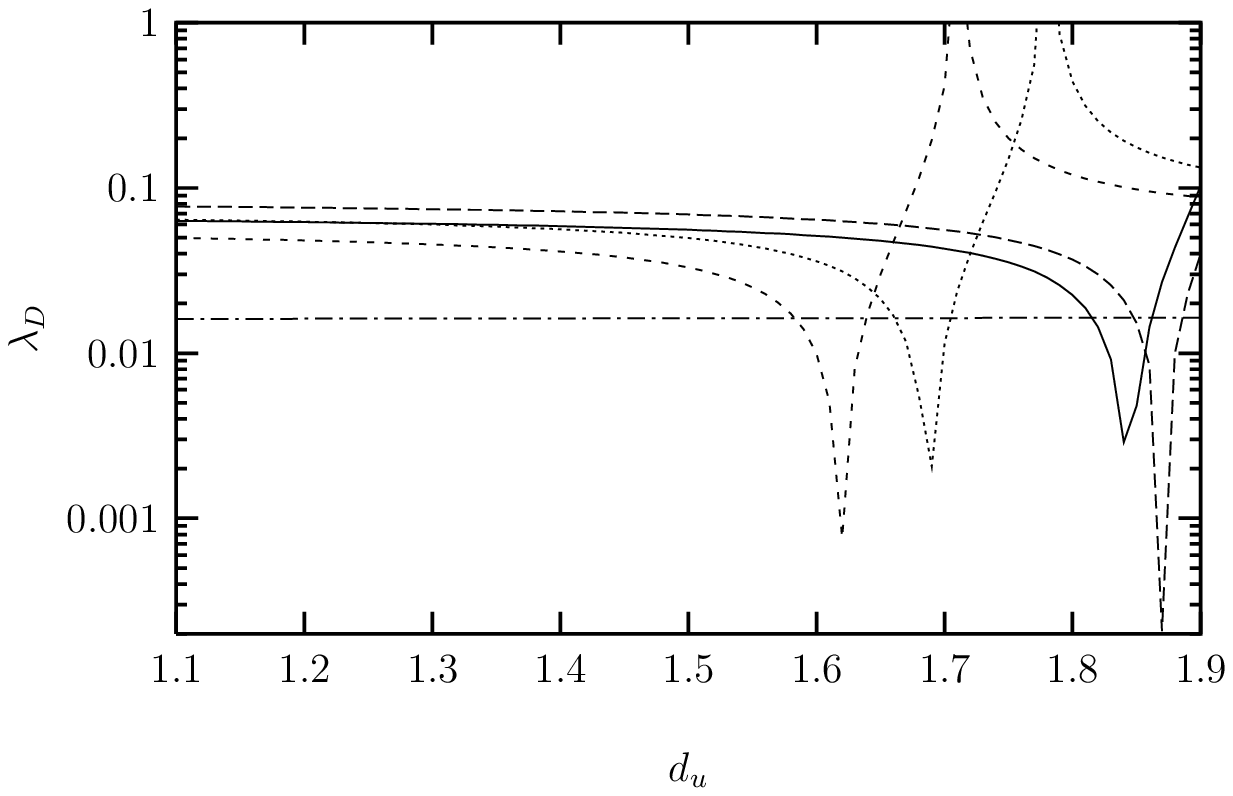} \vskip -3.0truein
\caption[]{$\lambda_D$ as a function of $d_u$ for $s_0=0.1$. Here
the solid-long dashed-dashed-dotted-dash dotted line represents
$\lambda_D$ for $m_I=110\,GeV$, $m_D=20\,GeV$-$m_I=120\,GeV$,
$m_D=20\,GeV$-$m_I=110\,GeV$, $m_D=30\,GeV$-$m_I=120\,GeV$,
$m_D=30\,GeV$-$m_I=110\,GeV$, $m_D=60\,GeV$. } \label{LambDdus001}
\end{figure}

In Fig.\ref{LambDdus001} we present $d_u$ dependence of
$\lambda_D$ for $s_0=0.1$. Here the solid-long
dashed-dashed-dotted-dash dotted line represents $\lambda_D$ for
$m_I=110\,GeV$, $m_D=20\,GeV$-$m_I=120\,GeV$,
$m_D=20\,GeV$-$m_I=110\,GeV$, $m_D=30\,GeV$-$m_I=120\,GeV$,
$m_D=30\,GeV$-$m_I=110\,GeV$, $m_D=60\,GeV$. It is observed that
for heavy Darkon $\lambda_D$ is not sensitive to $d_u$. For the
light Darkon particle case $\lambda_D$ is sensitive to $d_u$
around the numerical values which the threshold $m_D\sim
\frac{m_{II}}{2}$ are reached.

\begin{figure}[htb]
\vskip -3.0truein \centering \epsfxsize=6.8in
\leavevmode\epsffile{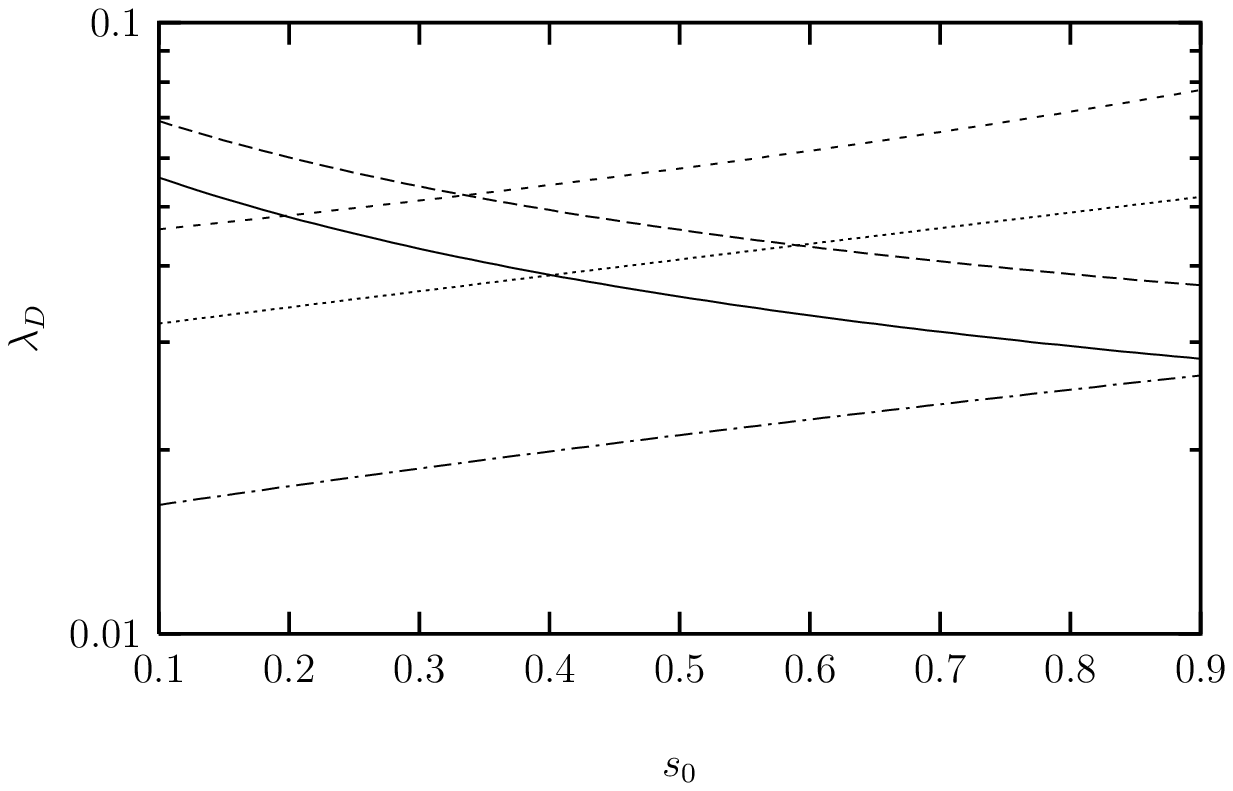} \vskip -3.0truein
\caption[]{$\lambda_D$ as a function of $s_0$ for $d_u=1.5$. Here
the solid-long dashed-dashed-dotted-dash dotted line represents
$\lambda_D$ for $m_I=110\,GeV$, $m_D=20\,GeV$-$m_I=120\,GeV$,
$m_D=20\,GeV$-$m_I=110\,GeV$, $m_D=70\,GeV$-$m_I=120\,GeV$,
$m_D=70\,GeV$-$m_I=110\,GeV$, $m_D=60\,GeV$. } \label{LambDs0du15}
\end{figure}

Fig.\ref{LambDs0du15} represents $s_0$ dependence of $\lambda_D$
for $d_u=1.5$. Here the solid-long dashed-dashed-dotted-dash
dotted line represents $\lambda_D$ for $m_I=110\,GeV$,
$m_D=20\,GeV$-$m_I=120\,GeV$, $m_D=20\,GeV$-$m_I=110\,GeV$,
$m_D=70\,GeV$-$m_I=120\,GeV$, $m_D=70\,GeV$-$m_I=110\,GeV$,
$m_D=60\,GeV$. This figure shows that for light (heavy) Darkon
$\lambda_D$ decreases (increases) with increasing values of $s_0$
since the increase in $s_0$ causes that the masses $m_{II}$ and
$m_{III}$ become lighter. With the decrease in mass $m_{II}$ $m_D$
reaches $\frac{m_{II}}{2}$ and the resonant annihilation occurs
for light Darkon. For the heavy one, $m_D$ becomes far from
$\frac{m_{II}}{2}$ as $s_0$ decreases and the annihilation rate
becomes small.

In summary, we consider that the electroweak symmetry breaking at
tree level occurs with the interaction of the SM Higgs doublet,
the hidden scalar and the hidden unparticle sector. Furthermore we
introduce an additional scalar SM singlet stable field $\phi_D$,
which is a dark matter candidate. This scalar disappears by
annihilating as pairs into SM particles with the help of the
exchange scalars, appearing after the electroweak symmetry
breaking and the mixing. By respecting the annihilation rate which
does not contradict with the current relic density, we predict the
tree level coupling $\lambda_D$ which drives the annihilation
process. We see that the unparticle scaling $d_u$ and the
parameter $s_0$ play considerable role on the annihilation and,
therefore, on the coupling $\lambda_D$. Once the dark matter mass
$m_D$ is fixed by the dark matter search experiments, it would be
possible to understand the mechanism behind the possible
annihilation process and one could get considerable information
about the electroweak symmetry breaking.
%%%
%%%
\newpage
{\Large \textbf{Appendix}}
\\ \\
Here we would like to present briefly (see \cite{Iltanmuonanogunp}
for details) the possible mechanism of the electroweak symmetry
breaking coming from the coupling of unparticle and the scalar
sector of the toy model used. The scalar potential which is
responsible for the unparticle-neutral scalars mixing reads:
\begin{eqnarray}
V(\Phi_1, \phi_2, \phi)&=& \lambda_0 (\Phi_1^\dagger\,
\Phi_1)^2+\lambda^\prime_0 (\phi_2^*\, \phi_2)^2+\lambda_1
(\phi^*\, \phi)^2 \nonumber \\ &+& 2 \lambda_2\,\mu^{2-d_u}\,
(\Phi_1^\dagger\, \Phi_1)\,(\phi^*\, \phi)^\frac{d_u}{2}+2
\lambda^\prime_2\,\mu^{2-d_u}\, (\phi_2^*\, \phi_2)\,(\phi^*\,
\phi)^\frac{d_u}{2}\, ,
 \label{potential}
\end{eqnarray}
where $\mu$ is the parameter inserted in order to make the
couplings $\lambda_2$ and $\lambda^\prime_2$ dimensionless. In
order to find the minimum of the potential V along the ray
$\Phi_{i}=\rho\, N_i$ with $\Phi_{i}=(\Phi_{1}, \phi_{2}, \phi)$
(see \cite{Weinberg})
\begin{eqnarray}
\Phi_{1}=\frac{\rho}{\sqrt{2}}\left(\begin{array}{c c}
0\\N_0\end{array}\right) \,\, ;
\phi_{2}=\frac{\rho}{\sqrt{2}}\,N^\prime_0 \,\,;
\phi=\frac{\rho}{\sqrt{2}}\,N_1 \, , \label{Phi12phi}
\end{eqnarray}
in unitary gauge, and the potential V
\begin{eqnarray}
V(\rho,N_i)&=& \frac{\rho^4}{4}\,\Bigg( \lambda_0\, N_0^4
+\lambda^\prime_0\, N^{\prime\, 4}_0 + 2\,\Big(
\frac{\hat{\rho}^2}{2}
\Big)^{-\epsilon}\,(\lambda_2\,N_0^2+\lambda^\prime_2\,N^{\prime\,
2}_0)\,N_1^{d_u}+\lambda_1\,N_1^4 \Bigg)\,, \label{potential2}
\end{eqnarray}
the stationary condition $\frac{\partial V}{
\partial N_i}|_{\vec{n}}$ along a special $\vec{n}$ direction
should be calculated\footnote{Here $\epsilon=\frac{2-d_u}{2}$ and
$\vec{N}$ is taken as the unit vector in the field space as
$N_0^2+N_0^{\prime 2}+N_1^2=1\, . $}.
Finally, one gets the minimum values of $N_i$, namely $n_i$ as
\begin{eqnarray}
n^2_0=\frac{\chi}{1+\chi+\kappa}\,\,, \,\,\, n^{\prime\,
^2}_0=\frac{1}{1+\chi+\kappa}\,\, , \,\,\,
n^2_1=\frac{\kappa}{1+\chi+\kappa}\,\, , \,\,\, \label{n01}
\end{eqnarray}
where
\begin{eqnarray}
\chi=\frac{\lambda^\prime_0}{\lambda_0}\,\,, \,\,\,
\kappa=\sqrt{\frac{d_u}{2}}\,\sqrt{\frac{\lambda^\prime_0\,
(\lambda_0+\lambda^\prime_0)} {\lambda_0\,\lambda_1}} \, ,
\label{chikappa2}
\end{eqnarray}
for $\lambda_2=\lambda^\prime_2$ which we consider in our
calculations. By using eq.(\ref{n01}), the nontrivial minimum
value of the potential is obtained as
\begin{eqnarray}
V(\rho, n_i)&=& -\frac{\rho^4}{4}\,\Bigg( \lambda_0\, n_0^4
+\lambda^\prime_0\, n^{\prime\, 4}_0  \Bigg)\,\epsilon\,.
\label{minpotential2}
\end{eqnarray}
This is the case that the minimum of the potential is nontrivial,
namely $V(\rho, n_i)\neq 0$ for $1< d_u <2$, without the need for
CW mechanism (see \cite{Weinberg} for details of CW mechanism).
The stationary condition fixes the parameter $\rho$ as,
\begin{eqnarray}
\rho=\rho_0=\Bigg
(\frac{-2^\epsilon\,\lambda_2\,n_1^d}{\lambda_0\,n_0^2}
 \Bigg)^\frac{1}{2\,\epsilon}\,\mu
\, ,\label{rho1a}
\end{eqnarray}
and one gets
\begin{eqnarray}
\hat{\rho_0}^2=(\frac{\rho_0}{\mu})^2=2\,\Big(\frac{d}{2}\Big)^{\frac{d}{4-2\,d}}\,
\Big(\frac{-\lambda_2}
{\lambda^\prime_0}\Big)\,\Big(1-\sqrt{\frac{d}{2}}\,\frac{\lambda^\prime_0}
{\lambda_2}+\frac{\lambda^\prime_0} {\lambda_0}\Big) \, ,
\label{rhorestr}
\end{eqnarray}
by using the eq.(\ref{n01}) with the help of the restriction
\begin{eqnarray}
\lambda_2=-\sqrt{\frac{\lambda_0\,\lambda^\prime_0\,\lambda_1}
{\lambda_0+\lambda^\prime_0 }} \, .\label{restr22}
\end{eqnarray}
Here the restriction in eq.(\ref{restr22}) arises when one chooses
$d_u=2$ in the stationary conditions.

Now, we  study the mixing matrix of the scalars under
consideration. The expansion of the fields $\Phi_{1}$, $\phi_{2}$
and $\phi$ around the vacuum
\begin{eqnarray}
\Phi_{1}=\frac{1}{\sqrt{2}}\left(\begin{array}{c c}
0\\\rho_0\,n_0+h\end{array}\right) \,\, ;
\phi_{2}=\frac{1}{\sqrt{2}}\,(\rho_0\,n^\prime_0+h^\prime) \,\,;
\phi=\frac{1}{\sqrt{2}}\,(\rho_0\,n_1+s) , \label{Phi12phi2}
\end{eqnarray}
results in the potential (eq. (\ref{potential}))
\begin{eqnarray}
V(h, h^\prime, s)\!\!\!\!\!&=&\!\!\!\!\! \frac{\lambda_0}{4}\,
(\rho_0\,n_0+h)^4+ \frac{\lambda^\prime_0}{4}\,
(\rho_0\,n^\prime_0+h^\prime)^4+\frac{\lambda_1}{4}\,
(\rho_0\,n_1+s)^4 + 2^{-\frac{d}{2}}\,
\lambda_2\,\mu^{2\,\epsilon}\,
(\rho_0\,n_0+h)^2\,(\rho_0\,n_1+s)^{d_u}\nonumber \\
&+&2^{-\frac{d}{2}}\, \lambda^\prime_2\,\mu^{2\,\epsilon}\,
(\rho_0\,n^\prime_0+h^\prime)^2\,(\rho_0\,n_1+s)^{d_u}\, ,
\label{potential3}
\end{eqnarray}
and the mass matrix
$(M^2)_{ij}=\frac{\partial^2\,V}{\partial\,\phi_i\,\partial\,\phi_j}|_{\phi_i=0}$
with $\phi_i=(h,h^\prime,s)$ as
\\
\begin{eqnarray}
(M^2)_{ij}=2\,\rho_0^2\,n_0^2\, \left(%
\begin{array}{ccc}
\lambda_0 & 0 & -\Big(\frac{d_u\,\lambda_0}{2} \Big)^\frac{3}{4}\,
\Big(\frac{\lambda^\prime_0\,\lambda_1}
{\lambda_0+\lambda^\prime_0} \Big)^\frac{1}{4} \\ \\0 & \lambda_0
& -\Big(\frac{d_u\,\lambda_0}{2} \Big)^\frac{3}{4}\,
\Big(\frac{\lambda_0^2\,\lambda_1}
{\lambda^\prime_0\,(\lambda_0+\lambda^\prime_0)} \Big)^\frac{1}{4}
\\ \\
 -\Big(\frac{d_u\,\lambda_0}{2} \Big)^\frac{3}{4}\,
\Big(\frac{\lambda^\prime_0\,\lambda_1}
{\lambda_0+\lambda^\prime_0} \Big)^\frac{1}{4} &
-\Big(\frac{d_u\,\lambda_0}{2}\Big)^\frac{3}{4}\,
\Big(\frac{\lambda_0^2\,\lambda_1}
{\lambda^\prime_0\,(\lambda_0+\lambda^\prime_0)} \Big)^\frac{1}{4}
& (2-\frac{d_u}{2})\,
\sqrt{\frac{d_u}{2}}\,\sqrt{\frac{\lambda_0\,\lambda_1\,
(\lambda_0+\lambda^\prime_0)}{\lambda^\prime_0}}\,, \\
\end{array}%
\right) \label{restr2}
\end{eqnarray}
with eigenvalues
\begin{eqnarray}
m_I^2&=& 2\,\lambda_0\,n_0^2\,\rho_0^2\, , \nonumber \\
m_{II}^2&=& \lambda_0\,n_0^2\,\rho_0^2\, \Bigg(
1+(2-\frac{d_u}{2})\,\sqrt{\frac{d_u\,s_{10}\,(1+s_0)}{2\,s_0}}-
\sqrt{\Delta}
\Bigg)\, , \nonumber \\
m_{III}^2&=& \lambda_0\,n_0^2\,\rho_0^2\, \Bigg(
1+(2-\frac{d_u}{2})\,\sqrt{\frac{d_u\,s_{10}\,(1+s_0)}{2\,s_0}}+
\sqrt{\Delta} \Bigg)\, ,\label{mass2}
\end{eqnarray}
where
\begin{eqnarray}
\Delta=d_u\,\sqrt{\frac{2 d_u\,s_{10}\,(1+s_0)}{s_0}}+\Bigg(
 1+(\frac{d_u}{2}-2)\,\sqrt{\frac{d_u\,s_{10}\,(1+s_0)}{2\,s_0}}\Bigg)^2
 \,.
 \label{delta}
\end{eqnarray}
Here we used the parametrization
\begin{eqnarray}
\lambda^\prime_0=s_0\,\lambda_0\,,\,\,\,
\lambda_1=s_{10}\,\lambda_0\  \, . \label{lambda01}
\end{eqnarray}
The physical states $h_I,\,h_{II},\,h_{III}$ are connected to the
original states $h,\,h^\prime,\,s$ as
\begin{eqnarray}
\left(%
\begin{array}{c}
  h \\
  h^\prime \\
  s \\
\end{array}%
\right)=\left(%
\begin{array}{ccc}
  c_\alpha & -c_\eta\,s_\alpha & s_\eta\,s_\alpha \\
  s_\alpha & c_\eta\,c_\alpha & -s_\eta\,c_\alpha \\
  0 & s_\eta & c_\eta \\
\end{array}%
\right)
\left(%
\begin{array}{c}
  h_I \\
  h_{II} \\
  h_{III} \\
\end{array}%
\right)\,, \label{diagMatrix}
\end{eqnarray}
where $c_{\alpha\,(\eta)}=cos\,\alpha\,(\eta)$,
$s_{\alpha\,(\eta)}=sin\,\alpha\,(\eta)$ and
\begin{eqnarray}
tan\,2\,\alpha&=& \frac{2\,\sqrt{s_0}}{s_0-1}\,, \nonumber
\\
tan\,2\,\eta&=& \Big(\frac{d_u}{2}\Big)^\frac{3}{4}\,\frac{2
\Big(s_0\,s_{10}\,(1+s_0)\,\Big)^\frac{1}{4}}{(1-\frac{d_u}{4})\,
\sqrt{2\,d_u\,s_{10}\,(1+s_0)}-\sqrt{s_0}}\, .
  \,\label{tan2alfeta}
\end{eqnarray}
When $d_u\rightarrow 2$, the state $h_{II}$ is massless in the
tree level and it has the lightest mass for $1< d_u< 2$. $h_{I}$
and $h_{III}$ can be identified as the SM Higgs boson and heavy
scalar coming from the shadow sector, respectively.

The final restriction is constructed by fixing the vacuum
expectation value $v_0=n_0\,\rho_0$, by the gauge boson mass $m_W$
as
\begin{eqnarray}
v_0^2=\frac{4\,m_W^2}{g_W^2}=\frac{1}{\sqrt{2}\,G_F}\, ,
\label{v02}
\end{eqnarray}
where $G_F$ is the Fermi constant. By using eqs. (\ref{n01}) and
(\ref{rhorestr}) we get
\begin{eqnarray}
\hat{v}_0^2=c_0\, \frac{s_{10}\,\sqrt{2\, s_0
\,(1+s_0)}+s_{0}\,\sqrt{d_u\,s_{10}}} {\sqrt{d\, s_0
\,(1+s_0)}+(1+s_0)\,\sqrt{2\,s_{10}}} \, , \label{hatv02}
\end{eqnarray}
with $c_0=2\,\Big(\frac{d_u}{2}\Big)^\frac{d_u}{2\,(2-\,d_u)}$.
The choice of the parameter $\mu$ around weak scale as $\mu=v_0$
results in the additional restriction which connects parameters
$s_0$ and $s_{10}$ (see eq. (\ref{hatv02}) by considering
$\hat{v}_0^2=1$) as
\begin{eqnarray}
s_{10}=\frac{1+s_0}{c_0^2\,s_0}  \, . \label{s10}
\end{eqnarray}
When $d_u\rightarrow 2$, $s_{10}\rightarrow
\frac{e}{4}\,\frac{1+s_0}{s_0}$ and when $d_u\rightarrow 1$,
$s_{10}\rightarrow \frac{1+s_0}{2\,s_0}$. It is shown that the
ratios are of the order of one and the choice $\mu=v_0$ is
reasonable (see \cite{JPLee} for the similar discussion.)
\newpage
\newpage
\end{document}